\documentclass[12pt]{article}
\usepackage[T2A]{fontenc}
\usepackage{graphicx}
\usepackage[cp1251]{inputenc}
\begin{document}
\newcommand{\iso}[2]{\mbox{$^{#1}{\rm #2}$}}
\newcommand{\Teff}{T_{\rm eff}}
\newcommand{\kms}{km\,s$^{-1}$}
\newcommand{\kH}{$S_{\!\rm H}$}    
\newcommand{\eps}[1]{\log\varepsilon_{\rm #1}}
\newcommand{\ba}[4]{\mbox{$#1 ^2{\rm #2}^{#3}_{\rm #4}$}}
\newcommand{\Vmac}{V_{\rm mac}}
\newcommand{\Eexc}{$E_{\rm exc}$}

\centerline{\bf Influence of Departures from LTE on Calcium, Titanium, and Iron}
\centerline{\bf Abundance Determinations in Cool Giants of Different Metallicities}

\bigskip
\centerline{Lyudmila Mashonkina\footnote{\tt E-mail: lima@inasan.ru}, Tatyana Sitnova, and Yuri Pakhomov}

\bigskip
\centerline{\it Institute of Astronomy, Russian Academy of Sciences,}
\centerline{\it Pyatnitskaya st. 48, 119017 Moscow, Russia}
 
\bigskip
{\bf Abstract -}
Non-local thermodynamic equilibrium (non-LTE) line formation for Ca~I-Ca~II, Ti~I-Ti~II, and Fe~I-Fe~II is considered in model atmospheres of giant stars with an effective temperature of 4000~K $\le \Teff \le$ 5000~K and a metal abundance of $-4 \le {\rm [Fe/H]} \le 0$. The departures from LTE are analyzed depending on atmospheric parameters. 
We present the non-LTE abundance corrections for 28 lines of Ca~I, 42 lines of Ti~I, 54 lines of Ti~II, and 262 lines of Fe~I 
and a three-dimensional interpolation code to obtain the non-LTE correction online ({\tt http://spectrum.inasan.ru/nLTE/}) for an individual spectral line and given atmospheric parameters.


\section{INTRODUCTION}

Chemical abundances of stars of different metallicity are used in studying the chemical evolution of galaxies, in particular, the Milky Way (MW) and dwarf satellite galaxies. We will use the iron abundance relative to the solar one, [Fe/H] = log~$(N_{\rm Fe}/N_{\rm H})_{star} - (N_{\rm Fe}/N_{\rm H})_\odot$, as a metallicity indicator.
The further into the low-metallicity region we want to advance, the more distant, on average,
objects should be observed. That is why the observational data for the [Fe/H] $< -3$ region in the MW are obtained mostly from giant stars, and only giants are accessible to high-resolution spectroscopy in distant
globular clusters and dwarf satellite galaxies. The spectral line formation conditions in the atmospheres
of, in particular, metal-deficient giants are far from the equilibrium ones. Nevertheless, in most cases,
not only the abundances of chemical elements but also the atmospheric parameters (effective temperature
$\Teff$, surface gravity log~g, and microturbulence $\xi_t$) are determined using the assumption of local thermodynamic equilibrium (LTE). For example, for the sample of stars from the classical paper by Cayrel~et~al. (2004), abundances of many elements were revised by Andrievsky et~al. (2007, Na;
2008, Al; 2009, Ba; 2010, Mg and K) and Spite et~al.(2012, Ca) based on the non-LTE line formation, 
but using, at the same time, log~g and [Fe/H] determined by Cayrel et~al. (2004) within LTE. As
shown by Mashonkina et~al. (2011) and Bergemann et~al. (2012), LTE underestimates the abundance
derived from Fe~I lines, and the effect increases with decreasing metallicity. This means that the ratio
[X(non-LTE)/Fe(LTE)] is overestimated by a larger amount at smaller [Fe/H], and we obtain a distorted
view of the change in the relative abundance X/Fe with metallicity (i.e., with time). Another source of
errors in determining X/Fe is the surface gravity obtained in LTE from the Fe~I/Fe~II ionization
equilibrium method. Since the Fe~I lines are subject to departures from LTE, while the non-LTE
effects for the Fe~II lines remain negligible down to a very low metallicity, [Fe/H] $\simeq -5$, LTE leads to
underestimated values of log~g. The LTE assumption is commonly used for stars in dwarf galaxies to determine
both atmospheric parameters (Frebel et~al. 2010; Simon et~al. 2010, 2015) and elemental abundances
(Tafelmeyer et~al. 2010; Gilmore et~al. 2013; Jablonka et~al. 2015). The only non-LTE paper is Sk{\'u}lad{\'o}ttir et~al. (2015), in which the sulfur abundance was determined for stars in the dwarf spheroidal (dSph) galaxy in Sculptor.

The goal of this paper is to study the systematic errors due to the use of the simplifying classical LTE
assumption in determining the abundances from lines of Ca~I, Ti~I, Ti~II, Fe~I, and Fe~II in cool giants in a wide metallicity range. We calculated the non-LTE abundance corrections for a large set of lines in the
ranges of stellar parameters 4000~K $\le \Teff \le$ 5000~K, $0.5 \le$ log~g $\le 2.5$, $-4 \le$ [Fe/H] $\le 0$ and treated a code for three-dimensional interpolation that allows the non-LTE correction for an individual spectral line and given atmospheric parameters to be obtained online. All the data are publicly available and can be used in the studies of red giants to determine their atmospheric parameters from Fe~I and Fe~II lines and the calcium, titanium, and iron abundances.

The non-LTE abundance corrections for Fe~I and Fe~II lines were computed in the literature (Lind et~al. 2012) in a wide range of stellar parameters, but this does not belittle the practical benefits of our work.
The point is that the non-LTE results for Fe~I-Fe~II and for other atoms depend strongly on a free parameter used in non-LTE calculations. This is the scaling factor \kH\ to the formulas of Steenbock and Holweger (1984) that were derived based on the classic theory of Drawin (1968) and are used to calculate the rate coefficients for excitation and ionization of atoms by inelastic collisions with neutral hydrogen atoms. This approach has been repeatedly criticized (see, e.g., Barklem et~al. 2011) for the groundlessness of applying the Drawin (1968) theory to the calculation of inelastic collisions with H~I atoms, however, we continue to use it in calculating the statistical equilibrium, because for most atoms there are neither laboratory measurements nor calculations of the cross sections for these processes. As shown by Mashonkina et~al. (2016), applying
the Al~I + H~I collision cross sections obtained in quantum-mechanical calculations gives an advantage
in analyzing the Al~I lines in stellar spectra compared to the formulas from Steenbock and Holweger
(1984), but, at the same time, if there are no accurate data, it is better to take into account, even if approximately, the inelastic collisions with H~I than to ignore them. Lind et~al. (2012) used the classic Drawin rates with \kH\ = 1 in their calculations. A different estimate (\kH\ = 0.5) was obtained by Sitnova et~al. (2015) when analyzing the Fe~I and Fe~II lines in a sample of dwarf stars in the range $-2.6 \le$ [Fe/H] $\le 0.2$. Our analysis of the iron lines in giants in the Sculptor dSph (see Section 3.3.1) confirmed the estimate 
by Sitnova et~al. (2015). Therefore, the non-LTE calculations for Fe~I-Fe~II were performed with \kH\ = 0.5.

The paper is structured as follows. The methods of calculations are described in Sect.~2. The departures from LTE for Ca~I, Ti~I-Ti~II, and Fe~I-Fe~II lines depending on atmospheric parameters are studied in Sect.~3. Section 4 provides the methodical recommendations and instructions for interpolation of the non-LTE corrections for given line and atmospheric parameters.

\section{THE METHODS OF CALCULATIONS}

We use the multilevel model atoms constructed using the most up-to-date atomic data. The techniques
of calculations were described in detail by Mashonkina et~al. (2007) for Ca~I-Ca~II, Sitnova et~al. (2016) for Ti~I-Ti~II, and Mashonkina et~al. (2011) for Fe~I-Fe~II. For Ca~I the calculation of inelastic collisions with hydrogen atoms was updated by using the results of quantum mechanical calculations from Belyaev et~al. (2016).

We solved the system of statistical equilibrium (SE) and radiative transfer equations in a given model atmosphere using the DETAIL code developed by Butler and Giddings (1985) based on the accelerated $\Lambda$-iteration method. The level populations obtained by solving the SE equations (non-LTE) and calculated from the Boltzmann-Saha formulas (LTE) were then used by the LINEC code (Sakhibullin 1983) to compute the theoretical non-LTE and LTE spectral line profiles, the equivalent widths $EW$, and the non-LTE abundance corrections
$\Delta_{\rm NLTE} = \eps{NLTE} - \eps{LTE}$. Here, $\eps{LTE}$ is the element abundance in the model atmosphere, and $\eps{NLTE}$ is the abundance at which the non-LTE calculations reproduce $EW$(LTE) of a given line.

The computations were performed for a grid of MARCS\footnote{\tt http://marcs.astro.uu.se} model atmospheres (Gustafsson et~al. 2008) with 4000~K $\le \Teff \le$ 5000~K, 0.5 $\le$ log~g $\le$ 2.5, and [Fe/H] = 0, $-1$, $-2$, $-2.5$, $-3$, $-3.5$, $-4$. The chemical composition is a standard one. This means that the relative abundance of the $\alpha$-process elements (O, Mg, Si, S, Ca) is [$\alpha$/Fe] = 0.4 in the metal-deficient models ([Fe/H] $\le -1$) and [$\alpha$/Fe] = 0 in the models with [Fe/H] = 0. The microturbulence is everywhere $\xi_t$ = 2\,\kms.

All the used model atmospheres were computed as spherically symmetric (SS) ones, but they are presented at the MARCS site in the form of plane-parallel (PP) models. This allowed them to be used with our codes, which solve the radiative transfer equation in PP geometry. For SS models with solar metal abundances, Heiter and Eriksson (2006) estimated, under the LTE assumption, the difference in the abundance deduced from a specific line between the self-consistent solution where the transfer equation is solved in SS geometry and the solution
in PP geometry. They concluded that the difference did not exceed 0.02~dex in absolute value if log~g $\ge 2$ and did not exceed 0.06~dex at the lower surface gravity (down to log~g = 0.5) for lines with $EW <$ 100~m\AA. Similar estimates were made by Pierre Norton (private communication) for five cool giants with
[Fe/H] $< -2$ from Jablonka et~al. (2015). For the 4540/1.15/$-2.45$ SS model, the self-consistent solution
leads to an increase in the abundance from Fe~I lines by 0.05~dex compared to the solution in PP geometry.
For the 4300/0.63/$-3.45$, 4400/1.01/$-3.20$, and 4600/1.19/$-3.30$ models, the difference did not
exceed 0.01~dex, but the reverse effect was detected for the 4480/1.01/$-3.88$ model with the greatest
metal deficiency: the abundance from Fe~I lines decreased by 0.12~dex. We cannot estimate the uncertainty
in our non-LTE results associated with the use of PP geometry for SS models and will assume that it
does not exceed the LTE estimates.

The list of lines includes 28/42/54/262/20 lines of Ca~I/Ti~I/Ti~II/Fe~I/Fe~II in the wavelength
range 3380-6745\,\AA, which are used by the Dwarf Abundances and Radial velocities Team (DART) in
analyzing high-resolution spectra (see, e.g., Tafelmeyer et~al. 2010, Table 5, and Jablonka et~al. 2015, Table 2).

\section{DEPARTURES FROM LTE FOR Ca~I, Ti~I-Ti~II, AND Fe~I-Fe~II LINES}

\begin{figure} 
  \resizebox{120mm}{!}{\includegraphics{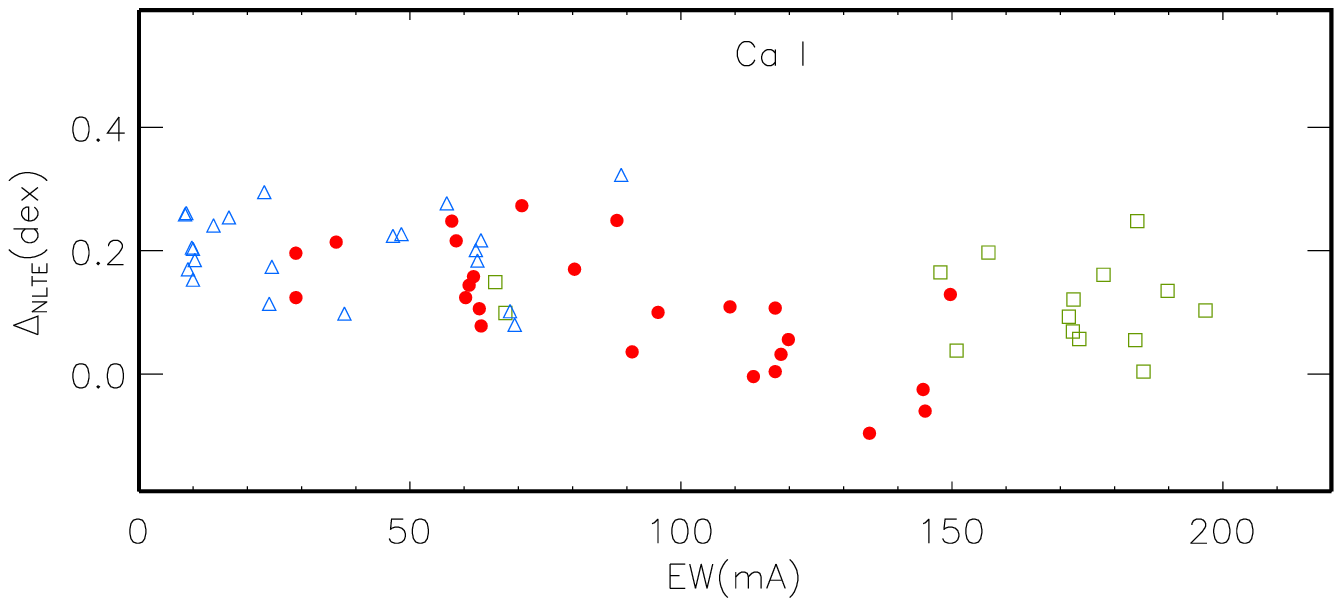}}
  \resizebox{120mm}{!}{\includegraphics{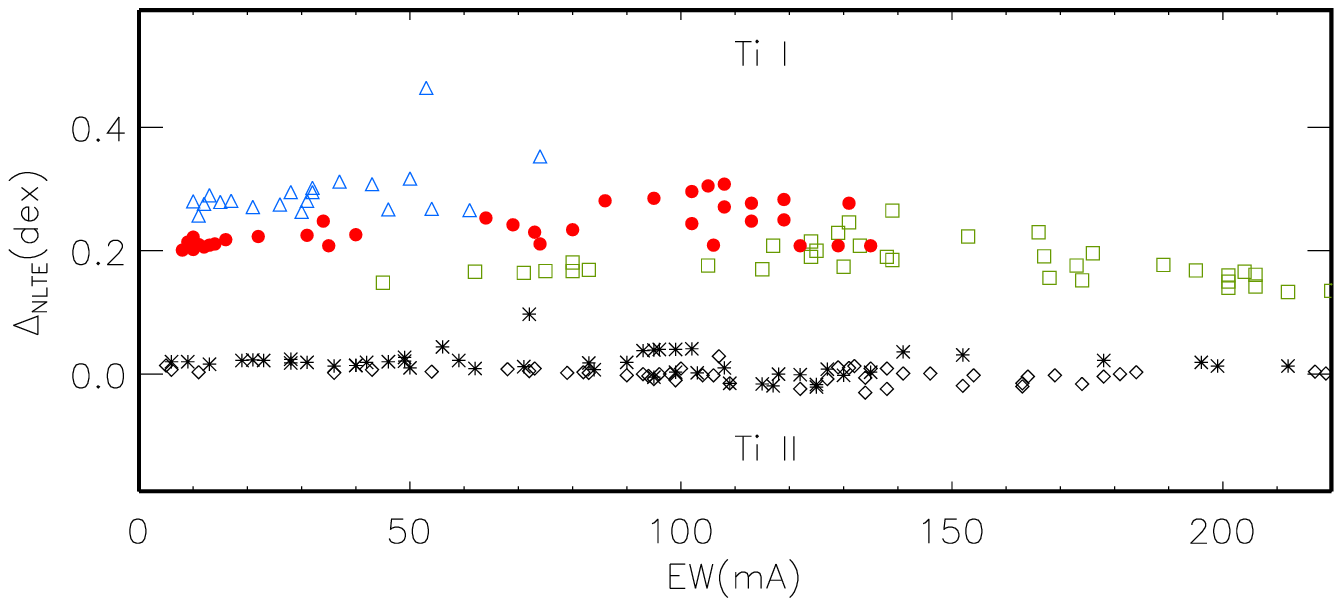}}
  \caption{Non-LTE corrections for lines of Ca~I (top panel) and Ti~I and Ti~II (bottom panel) as a function of equivalent width $EW$ in models with different metal abundances. The designation [M/H] is used instead of [Fe/H] everywhere in the figures. Lines of Ca~I and Ti~I are shown by the squares ([M/H] = 0), circles ([M/H] = $-2$), and triangles ([M/H] = $-3$). The Ti~II lines are shown by the diamonds ([M/H] = $-2$) and asterisks ([M/H] = $-3$). Everywhere, $\Teff$ = 4500~K and log~g = 1.0. The lines with $EW >$ 200~m\AA\ are not shown.}
\label{Fig:ca_ti_metallicity}
\end{figure}

\begin{figure} 
  \resizebox{120mm}{!}{\includegraphics{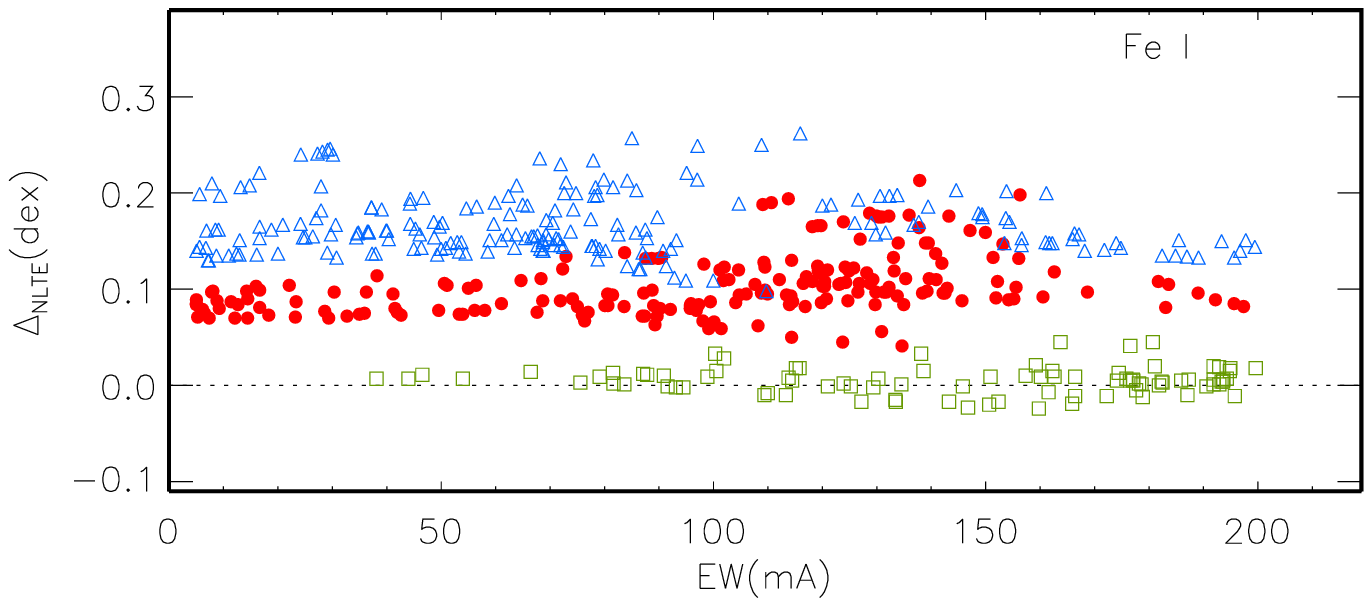}}
  \resizebox{120mm}{!}{\includegraphics{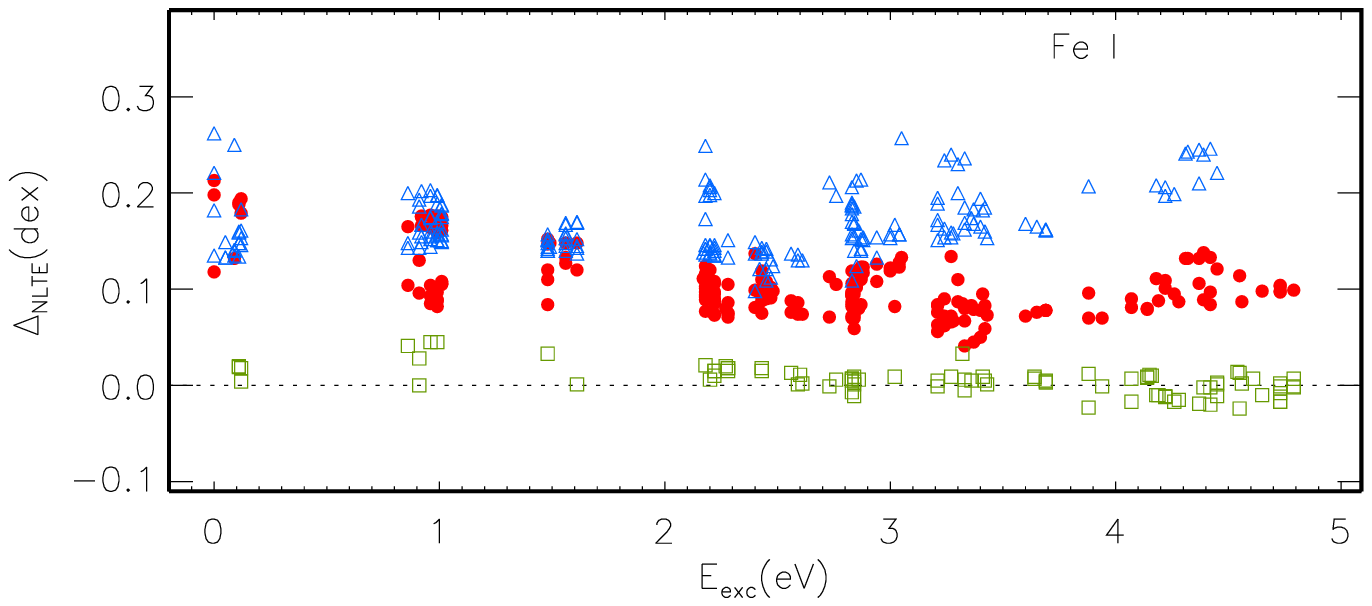}}
\caption{Same as in Fig.\,\ref{Fig:ca_ti_metallicity} for lines of Fe~I as a function of $EW$ (top panel) and \Eexc\ (bottom panel). Everywhere, \kH\ = 0.5.}
\label{Fig:fe_dnlte_metallicity}
\end{figure}

Since calcium, titanium, and iron are highly ionized in stellar atmospheres with $\Teff$/log~g in the range
under consideration, the number densities of neutral atoms (Ca~I, Ti~I, and Fe~I) easily deviate from their
equilibrium values when the mean intensity of ionizing radiation, $J_\nu$, deviates from the Planck function
$B_\nu(T)$. As was discussed in our previous papers and by our predecessors (for references, see Mashonkina
et~al. (2007) for Ca~I, Sitnova et~al. (2016) for Ti~I, and Mashonkina et~al. (2011) for Fe~I), an excess of
$J_\nu$ over $B_\nu(T)$ in the ultraviolet (UV) range leads to overionization of these atoms, i.e., to a decrease in the level populations compared to the equilibrium ones in the atmospheric layers where the medium is optically transparent to radiation beyond the ionization threshold for low-excitation levels. Therefore,
the Ca~I, Ti~I, and Fe~I lines are weakened in the non-LTE calculations compared with their LTE strengths, and the non-LTE abundance corrections are positive. The non-LTE effects are enhanced with decreasing metal
abundance, because the collisional processes become progressively less efficient due to the decrease in electron number density, while the radiative processes, on the contrary, become increasingly efficient due to the
decrease in opacity in the UV. Figures\,\ref{Fig:ca_ti_metallicity} and \ref{Fig:fe_dnlte_metallicity} serve as an illustration.

The number densities of Ti~II and Fe~II ions remain the equilibrium ones throughout the atmosphere, however, the populations of excited levels can depart from LTE due to
the UV pumping transitions from the ground and low-excitation states. In each model atmosphere,
the non-LTE effects in lines of Ti~II (Fig.\,\ref{Fig:ca_ti_metallicity}) and Fe~II are much smaller than those in Ca~I, Ti~I, and Fe~I lines.

\begin{figure} 
  \resizebox{120mm}{!}{\includegraphics{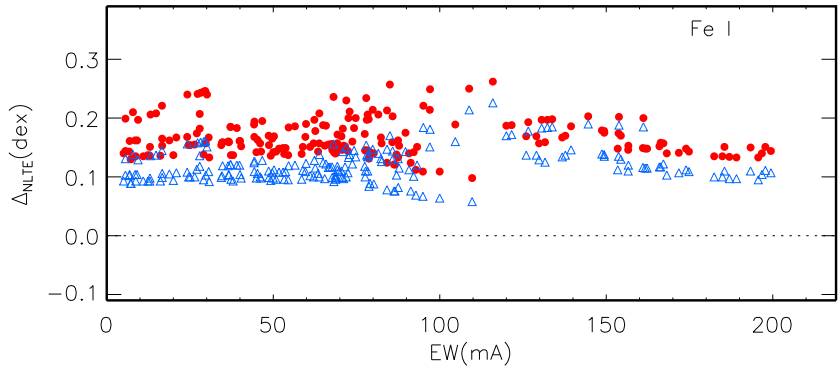}}
  \caption{Influence of the choice of \kH\ on the abundance determination from iron lines: the non-LTE corrections for lines of Fe~I in the 4500/1.0/$-3$ model at \kH\ = 1 (triangles) and 0.5 (circles). }
\label{Fig:sh} 
\end{figure}

The main source of uncertainties in our non-LTE results for Ti~I-Ti~II and Fe~I-Fe~II is an approximate treatment of inelastic collisions with H~I in the SE calculations due to the absence of accurate data on the cross sections for the processes. We use the classic Drawinian rates with a scaling factor \kH. For Ti~I-Ti~II and Fe~I-Fe~II, estimates of \kH\ = 1 and 0.5, respectively, were made by Sitnova et~al. (2016, 2015) through the analysis of lines in two ionization
stages for a sample of dwarf stars with [Fe/H] $< -1$. As can be seen from Fig.\,\ref{Fig:sh}, the choice of \kH\ affects significantly the non-LTE correction for the Fe~I lines in the $\Teff$/log~g/[Fe/H] = 4500/1/$-3$ model. Therefore, the values of \kH\ for the titanium and iron lines
were checked using giant stars with [Fe/H] $< -2$.

\subsection{Ti~I/Ti~II and Fe~I/Fe~II Ionization Equilibrium in the Atmospheres of Metal-Poor Giants in the Sculptor dSph}

\begin{table} 
 \caption{\label{Tab:parameters} Atmospheric parameters of the investigated stars in the Sculptor dSph galaxy. The microturbulence is given in \kms. }
 \centering
 \begin{tabular}{lccccc}
\hline\hline \noalign{\smallskip}
Star & $\Teff$ (K) &  log~g & Ref. &   [Fe/H] &  $\xi_t$ \\
 \noalign{\smallskip} \hline \noalign{\smallskip}
ET0381     &  4570 &  1.17 & JNM2015 & $-$2.16 & 1.7  \\
Scl002\_06  &  4390 &  0.68 & JNM2015 & $-$3.15 & 2.3  \\
Scl03\_059  &  4530 &  1.08 & JNM2015 & $-$2.88 & 1.9  \\
Scl031\_11  &  4670 &  1.13 & JNM2015 & $-$3.69 & 2.0 \\
Scl074\_02  &  4680 &  1.23 & JNM2015 & $-$3.06 & 2.0  \\
Scl07-49   &  4630 &  1.28 & TJH2010 & $-$2.99 & 2.4  \\
Scl07-50   &  4800 &  1.56 & TJH2010 & $-$4.00 & 2.2  \\
\noalign{\smallskip}\hline \noalign{\smallskip}
\multicolumn{6}{l}{JNM2015 = Jablonka et~al. (2015), } \\
\multicolumn{6}{l}{TJH2010 = Tafelmeyer et~al. (2010) } \\
\end{tabular}
\end{table}

Seven very metal-poor (VMP) giants in the Sculptor dSph were taken from Tafelmeyer et~al. (2010) and Jablonka et~al. (2015). The stars in a galaxy have the advantage that they are at the
same known distance. This allows the surface gravity to be calculated if the effective temperature has been determined and the stellar mass is known. Since the stars are old, $M = 0.8 M_\odot$ is a reasonable constraint on their mass. The photometric temperatures from the V-I, V-J, and V-K colors and log~g were determined in the cited papers. We deduced the iron abundance from lines of Fe~II and $\xi_t$ from lines of Fe~I through a non-LTE analysis. More details will be given by Mashonkina (2016, in preparation). Here, we note that mostly experimental oscillator strengths were used for  lines of Fe~I, as given in the Vienna Atomic Line Database (VALD, Ryabchikova et~al. 2015). For  lines of Fe~II their log~gf calculated by Raassen and Uylings (1998) were corrected by a factor of $\Delta$log~$gf$ = +0.11 as recommended by Grevesse and Sauval (1999). The van der Waals
broadening constants for most lines were calculated
using the modern perturbation theory (Barklem et~al. 2000; Barklem and Aspelund-Johansson 2005).
Stellar parameters are given in Table\,\ref{Tab:parameters}.

\begin{figure} 
  \resizebox{120mm}{!}{\includegraphics{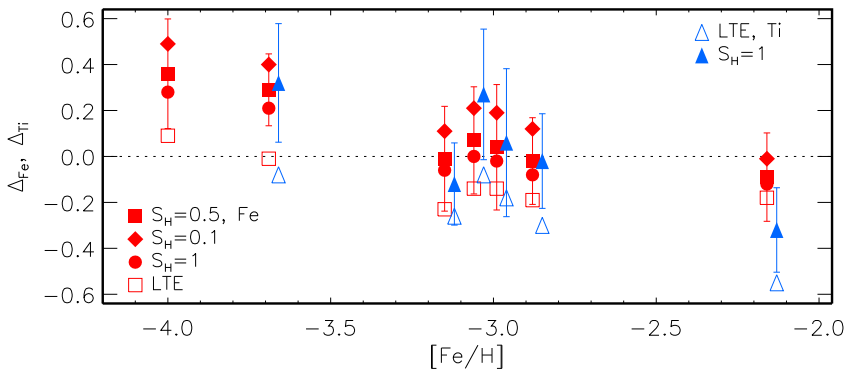}}
  \caption{Differences in the abundances derived from the lines of two ionization stages for iron and titanium for stars in the Sculptor galaxy in different spectral line formation scenarios.
The LTE calculations are indicated by the open squares (iron) and triangles (titanium).
The filled symbols correspond to the non-LTE calculations for Fe~I-Fe~II with \kH\ = 0.1 (diamonds), 0.5 (squares), 1 (circles) and Ti~I-Ti~II with \kH\ = 1 (triangles). To avoid the image superposition, the symbols for titanium are displaced along the horizontal axis by +0.03~dex. }
\label{Fig:fe1_fe2} 
\end{figure}

We use the published equivalent widths measured by Tafelmeyer et~al. (2010) and Jablonka et~al. (2015) in high-resolution spectra with R = $\lambda/\Delta\lambda$ = 45 000. The titanium and iron abundances were determined separately from the lines of each ionization stage
in different line formation scenarios, namely LTE and non-LTE with different \kH\. The abundance
differences between the two ionization stages, $\Delta_{\rm Fe}$ = $\eps{FeI}$ -- $\eps{FeII}$ and $\Delta_{\rm Ti}$ = $\eps{TiI}$ -- $\eps{TiII}$, are shown in Fig.\,\ref{Fig:fe1_fe2}. To avoid the image superposition, the errors in the abundance difference are shown for
one scenario. They were obtained by the quadratic sum of the errors of both ionisation stages,
$\sigma_\Delta = \sqrt{\sigma_{\rm I}^2 + \sigma_{\rm II}^2}$.
In turn, the statistical error of each ionisation stage is the dispersion in 
the single line measurements about the mean: 
$\sigma_{\rm I} = \sqrt{\Sigma(\overline{x}-x_i)^2/(N_{\rm I}-1)}$, where $N_{\rm I}$ is
the number of measured lines in the ionization stage I. Note that both $\sigma_{\rm I}$ and $\sigma_\Delta$ turn out to be large for all stars, at the order of 0.2~dex for both iron and titanium, probably because of the low signal-to-noise ratio, S/N = 15 at $\lambda$ = 4500\,\AA\ and increases to 70 at $\lambda$ = 6800\,\AA.

Let us consider separately five stars with [Fe/H] $> -3.2$ and two most MP stars, with [Fe/H] $< -3.6$. For the first group, our LTE analysis gives a lower abundance from the lines of neutral
atoms than that from the lines of ions, and the difference reaches $\Delta_{\rm Fe} = -0.23$~dex and $\Delta_{\rm Ti} = -0.55$~dex. In contrast, our non-LTE calculations with \kH\ = 0.1 lead to a higher abundance from the Fe~I lines than that from the Fe~II lines. An exception is the star ET0381 with the highest [Fe/H], for which precisely \kH\ = 0.1 leads to excellent agreement between the abundances from the lines of two ionization stages. For the other four objects, an optimal choice is \kH\ = 1 for Ti~I-Ti~II and \kH\ = 0.5 for Fe~I-Fe~II. However, it should be noted that $\Delta_{\rm Fe}$ does not exceed the determination error in the case of \kH\ = 1
either. For Ti~I/Ti~II in ET0381, it would be better to take \kH\ $< 1$.

For both stars with [Fe/H] $< -3.6$, consistent abundances from Fe~I and Fe~II are achieved at LTE, just as from the Ti~I and Ti~II lines for Scl031\_11. The Ti~I lines cannot be measured in
the spectrum of Scl07-50. The non-LTE abundances from the Fe~I and Ti~I lines are higher than those from the Fe~II and Ti~II lines. The reasons can be as follows.

(1) Poor statistics of lines. At such a low metallicity, the Ti~I lines are either very weak or disappear, and only two Fe~II lines, 4923\,\AA\ and 5018\,\AA, are observed in
the visible range. There are large uncertainties in gf for both. For example, log~gf for Fe~II 4923\,\AA\ varies in different papers between $-1.21$ (Schnabel et~al. 2004)
and $-1.50$ (Raassen and Uylings 1998).

(2) Uncertainties in the effective temperature. Since the calibrations from Ramirez and Melendez
(2005) used to determine $\Teff$ were deduced for [Fe/H] $\ge -3$, the errors for extremely metal-poor stars can exceed 100~K. With the temperatures revised downward by 170~K and 200~K for Scl031\_11 and Scl07-50, respectively, the non-LTE abundances from the
lines of two ionization stages could be reconciled for both titanium and iron.

(3) Uncertainties in treatment of inelastic collisions with H~I atoms. We realize that applying a
common scaling factor to the Drawinian rates for each transition in an atom is a very rough approximation. Comparison of the exact rates with the Drawinian ones for Mg~I shows that their ratio can be both much smaller and much greater than unity (Barklem et~al. 2012). An empirical estimate of \kH\ was made, using stars with [Fe/H] $> -3.2$, but the formation depths for radiation in the lines and below the ionization threshold change with metallicity, and the
role of different transitions in establishing the SE of an atom changes. For example, the overionization of Fe~I is caused by enhanced photoionization of levels with an excitation energy of \Eexc\ = 3-4.5~eV for moderately MP models ([Fe/H] $> -1$), while the levels with a lower energy, down to \Eexc\ =  1.4~eV, play a significant role in models with [Fe/H] $\simeq -2$. Accurate
data for inelastic Fe~I + H~I collisions are highly desirable to understand why non-LTE 'works poorly' at low metallicities.

Thus, our analysis of the Fe~I/Fe~II and Ti~I/Ti~II ionization equilibrium for giants with $-3.2 <$ [Fe/H] $< -2$ confirms the estimates of \kH\ = 0.5 and 1 obtained from analysis of the dwarf stars. Exactly these values were used in computing the grids of non-LTE corrections.

\subsection{The Non-LTE Corrections Depending on Atmospheric Parameters}

We describe the non-LTE effects depending on atmospheric parameters separately for the
Ti~I and Fe~I lines showing a common behavior, the Ca~I lines for which not only the magnitude but also the sign of the non-LTE correction can be different in the same model atmosphere, and the Ti~II and Fe~II lines of the dominant ionization stages.

We begin with the simplest case, the Ti~II and Fe~II lines, for which the departures from LTE are
small in the entire range of stellar parameters under consideration. For each of the 20 Fe~II lines, $\Delta_{\rm NLTE}$ does not exceed 0.01~dex in absolute value in all
models with $\Teff \le$ 4500~K and 0.02~dex in the models with $\Teff$ = 4750 and 5000~K, log~g $\ge 1$. For the two highest temperatures and log~g = 0.5, the correction
is still small in the models with [Fe/H] $\ge -1$ and increases from $\Delta_{\rm NLTE} = -0.03$~dex at [Fe/H] = $-2$ to $\Delta_{\rm NLTE} = +0.04$~dex at [Fe/H] = $-4$. Since the
quantities are small, we do not publish the table of non-LTE corrections for the Fe~II lines.

\begin{figure}  
\resizebox{50mm}{!}{\includegraphics{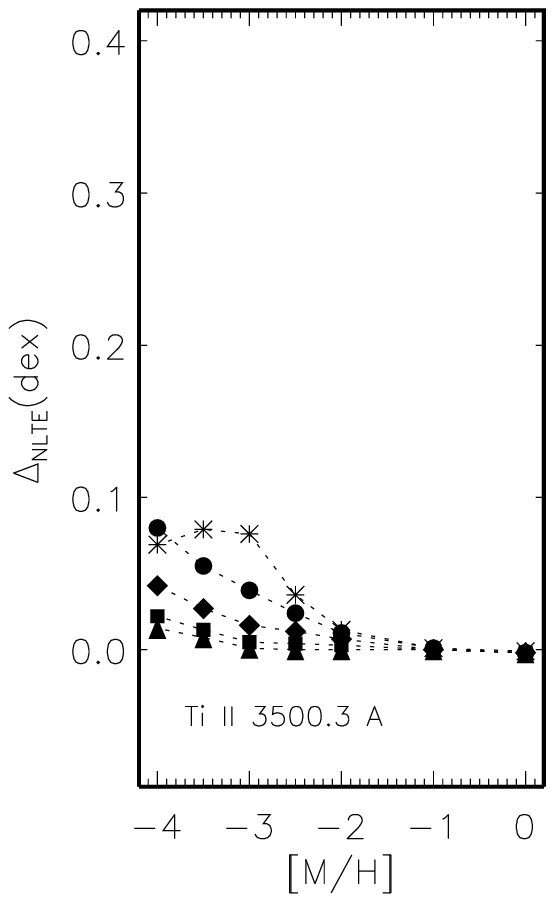}}
\resizebox{50mm}{!}{\includegraphics{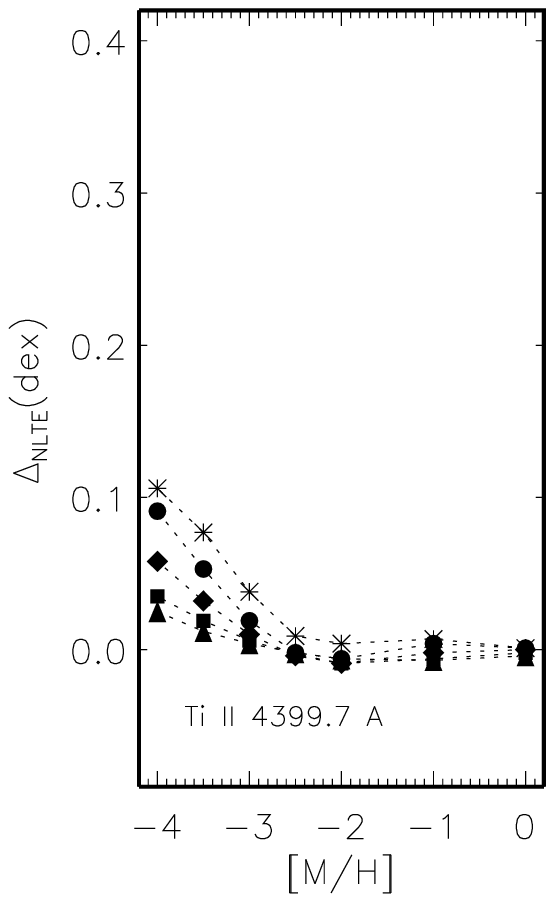}}
\resizebox{50mm}{!}{\includegraphics{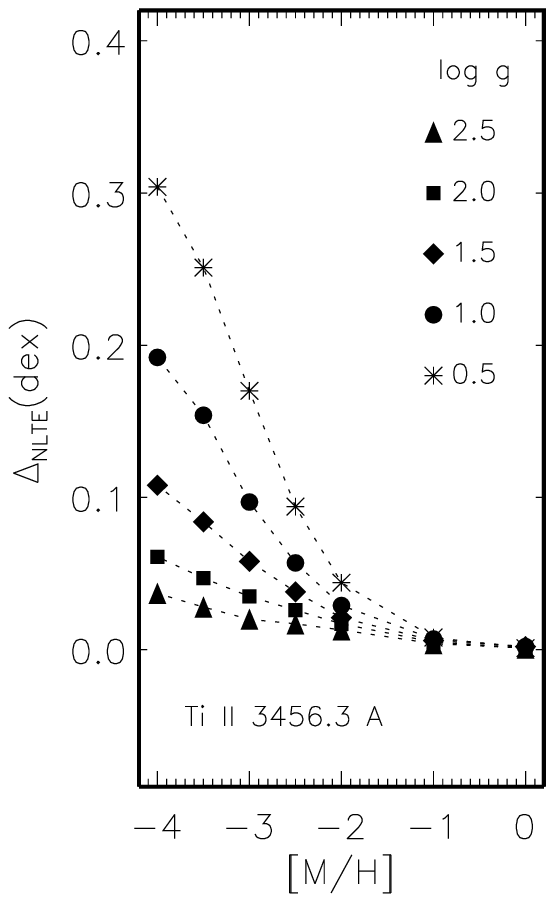}}
\caption{Non-LTE corrections for the selected lines of Ti~II in models with different metal abundances and surface gravities: log~g = 0.5 (asterisks), circles), 1.5 (diamonds), 2.0 (squares), and 2.5 (triangles). Everywhere, $\Teff$ = 4500~K.}
\label{Fig:ti2_lines}
\end{figure}

The non-LTE corrections for the Ti~II lines are larger than those for the Fe~II lines in the same
model. Figure\,\ref{Fig:ca_ti_metallicity} shows the non-LTE corrections
for all Ti~II lines in the 4500/1/$-2$ and 4500/1/$-3$ models, while Fig.\,\ref{Fig:ti2_lines} shows the same for three lines with different excitation energy of the lower level, \Eexc\ =
0.12~eV (3500\,\AA), 1.24~eV (4399\,\AA), and 2.06~eV (3456\,\AA), as a function of [Fe/H] and log~g at $\Teff$ = 4500~K. The departures from LTE are negligible in the models with [Fe/H] $\ge -1$. At lower metallicity, the corrections are everywhere positive, and their value increases with decreasing log~g and decreasing [Fe/H]. For low-excitation lines, $\Delta_{\rm NLTE}$ nowhere exceeds 0.1~dex, but the corrections are significant for Ti~II 3456\,\AA.

\begin{figure}  
\resizebox{50mm}{!}{\includegraphics{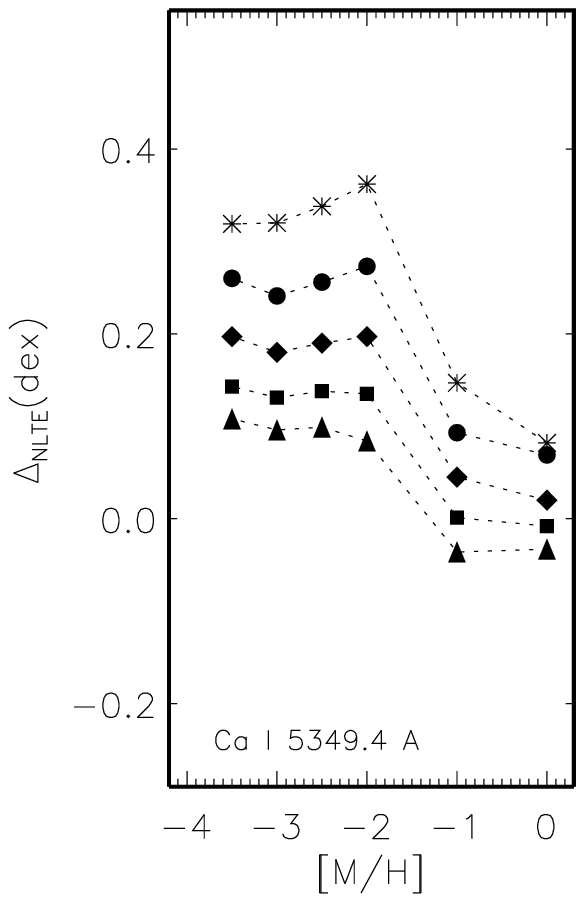}}
\resizebox{50mm}{!}{\includegraphics{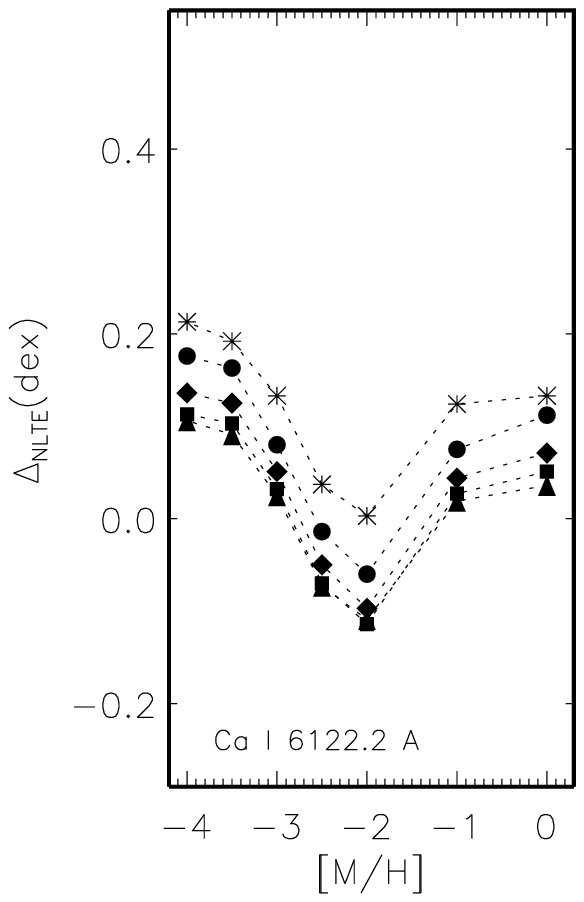}}
\resizebox{50mm}{!}{\includegraphics{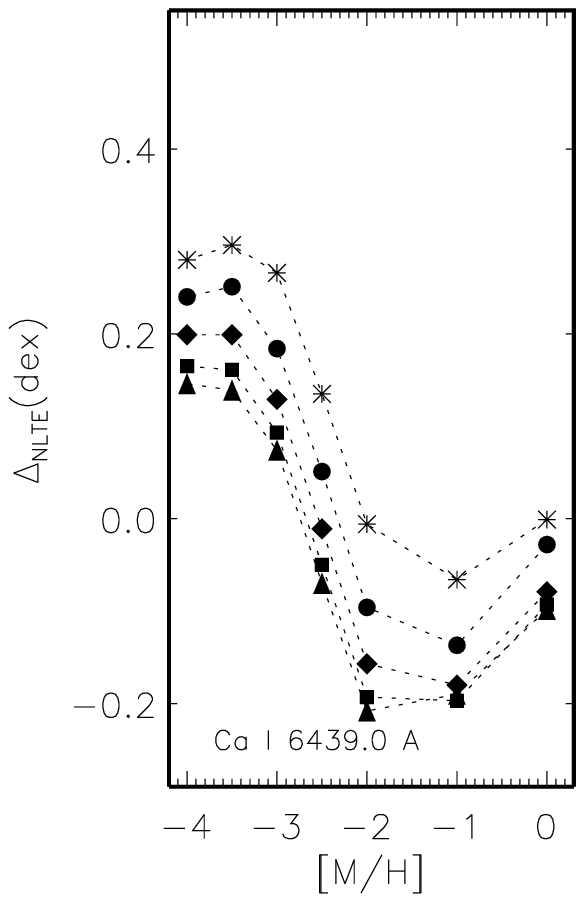}}
\resizebox{50mm}{!}{\includegraphics{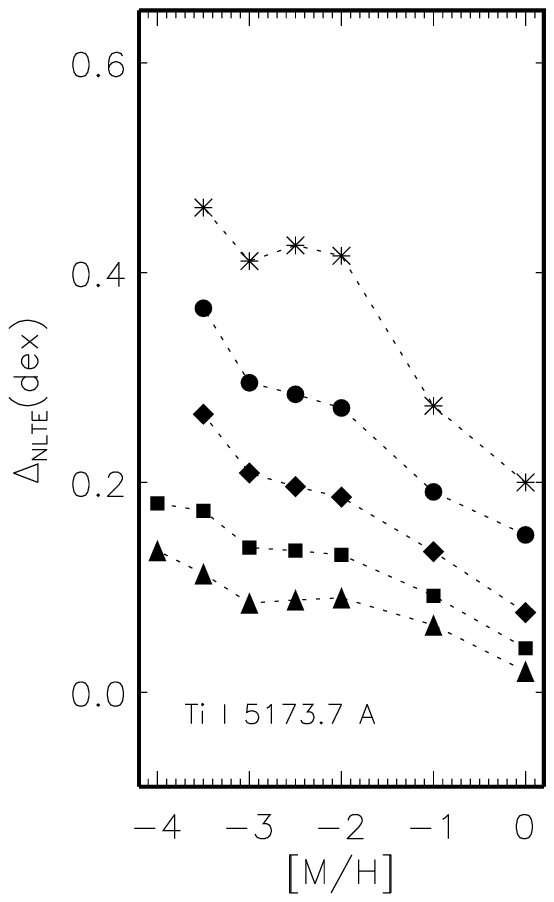}}
\resizebox{50mm}{!}{\includegraphics{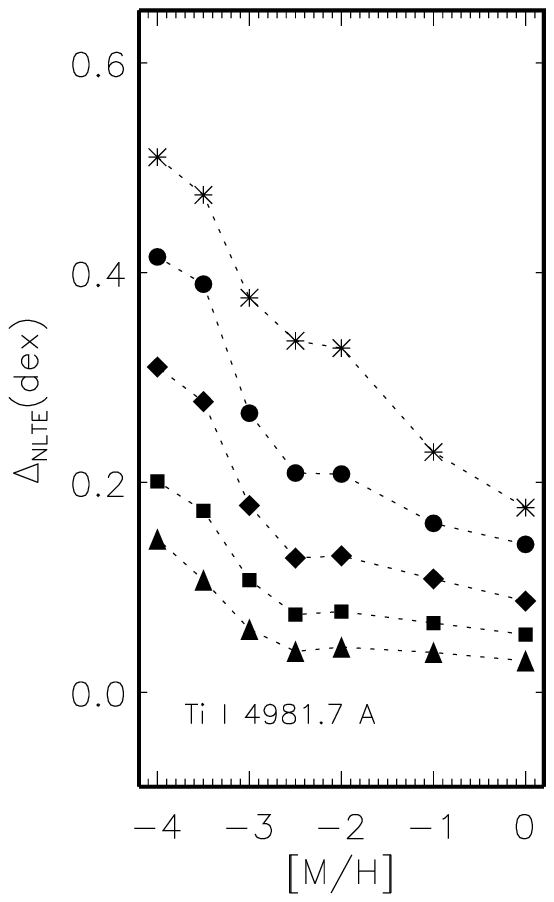}}
\resizebox{50mm}{!}{\includegraphics{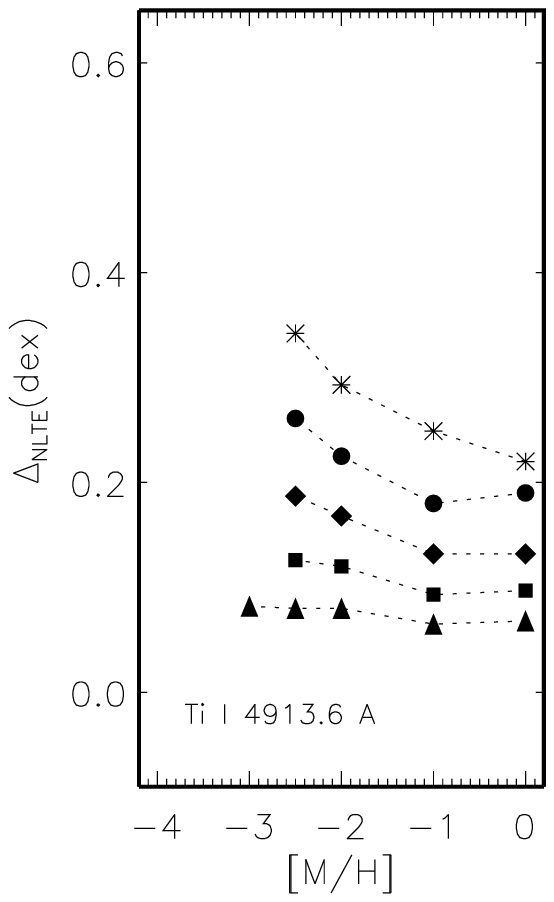}}
\resizebox{50mm}{!}{\includegraphics{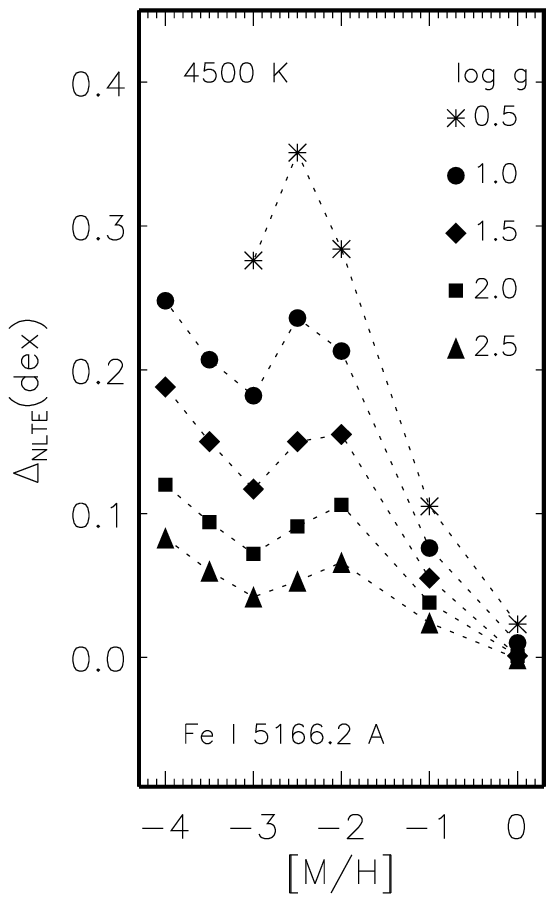}}
\resizebox{50mm}{!}{\includegraphics{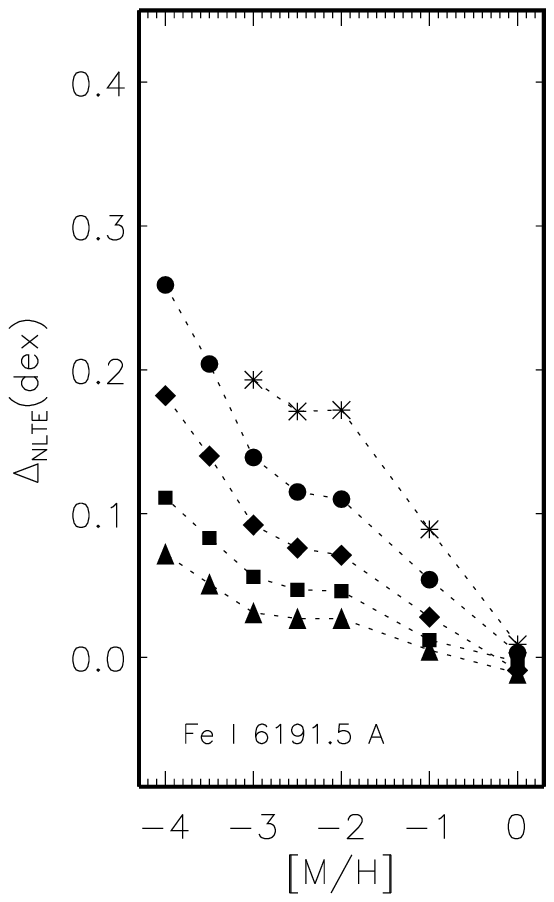}}
\resizebox{50mm}{!}{\includegraphics{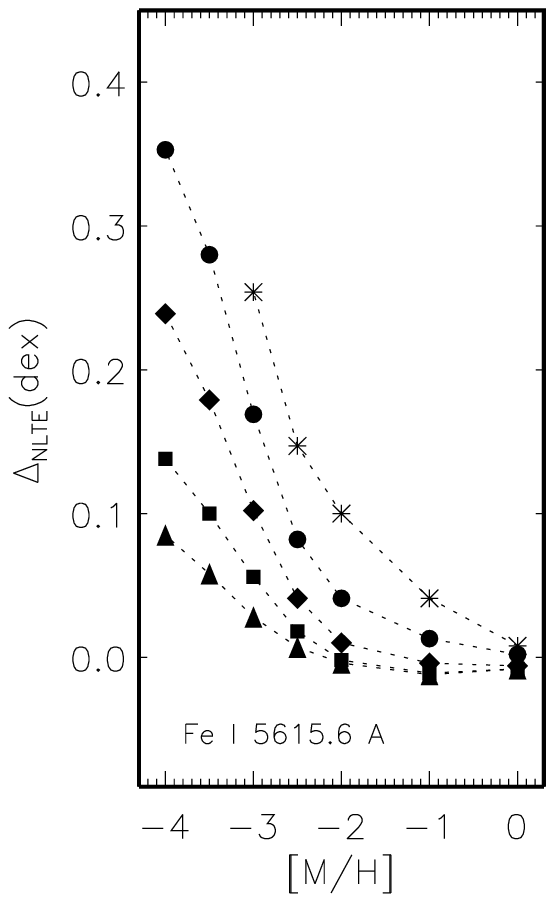}}
\caption{Same as in Fig.\,\ref{Fig:ti2_lines} for lines of Ca~I (top row), Ti~I (middle row), and Fe~I (bottom row). The non-LTE corrections are not provided for lines with $EW$(LTE) $< 3$~m\AA.}
\label{fig:lines}
\end{figure}

The dominant mechanism of departures from LTE for Ti~I and Fe~I is UV overionization, which leads
to a weakening of the lines in the entire range of stellar parameters under consideration. As expected, the non-LTE abundance corrections are positive for all of the calculated Ti~I and Fe~I lines and increase with decreasing [Fe/H] and decreasing log~g (Figs.\,\ref{Fig:ca_ti_metallicity}, \ref{Fig:fe_dnlte_metallicity}, and \ref{fig:lines}). In each model the non-LTE effects for Fe~I are smaller than those for Ti~I. This manifests itself particularly clearly in the models with [Fe/H] $\ge -1$. For the Fe~I lines the correction is close to zero at
[Fe/H] = 0 irrespective of log~g and does not exceed 0.1~dex at [Fe/H] = $-1$ and log~g $\ge 1$, while for the Ti~I lines it can reach 0.2~dex even at solar metallicity.
This is because the statistical equilibrium of Fe~I in this range of parameters is determined not only by overionization but also by a bulk of the UV transitions, in which detailed balance is
retained up to the outermost atmospheric layers as long as the medium remains opaque in the corresponding UV lines. In contrast, all transitions in Ti~I are weaker because of the lower abundance and cannot compete with overionization.

Figure\,\ref{Fig:fe_dnlte_metallicity} shows that $\Delta_{\rm NLTE}$ depends weakly on the
line equivalent width but depends on \Eexc\ in MP models, namely the corrections are larger for
low-excitation lines. This agrees qualitatively with the results for Fe~I lines in dwarf models (Mashonkina et~al. 2011; Lind et~al. 2012). In the 4500/1/$-2$ model the correction retains, on average, its value for Fe~I with \Eexc\ $> 2$~eV, but $\Delta_{\rm NLTE}$ begins to increase at
\Eexc\ $> 3$~eV in the model with [Fe/H] = $-3$.

\begin{figure}  
\resizebox{150mm}{!}{\includegraphics{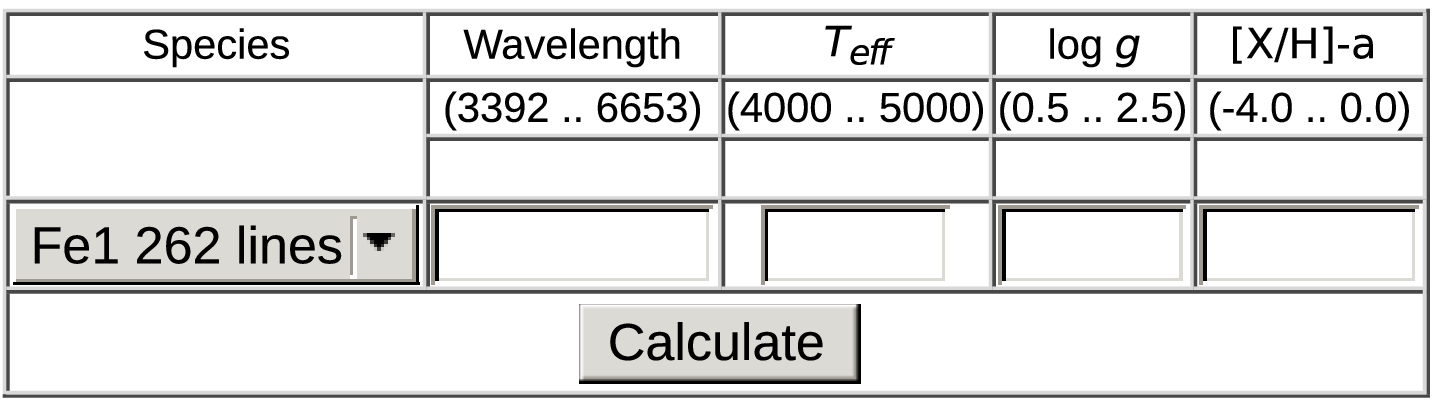}}
\caption{Query form for interpolating the non-LTE corrections.}
\label{Fig:online}
\end{figure}

For Ca~I the dominant mechanism of departures from LTE is also the UV overionization. Therefore, non-LTE leads to a weakening of the lines forming in deep layers. However, as was discussed in detail by Mashonkina et~al. (2007), another mechanism is at work for strong lines. This is the upper level depopulation in spontaneous transitions and dropping the line source function below the Planck function in the layers where the line core is formed, resulting in a strengthening of the line core. The combined effect depends on the relative contribution of the wings and the core to the total line absorption. For example, in any model the Ca~I 6439\,\AA\ line has weaker wings than does Ca~I 6122\,\AA\ due to the smaller (by 1.28~dex!) van der Waals broadening constant $C_6$. Our calculations show that, in the models with [Fe/H] = 0 and $-1$, the line wings weakened by overionization dominate in the total absorption of Ca~I 6122\,\AA, resulting in positive $\Delta_{\rm NLTE}$ irrespective of $\Teff$ and log~g (Fig.\,\ref{fig:lines}). A decrease in the contribution of the line wings to the total Ca~I 6122\,\AA\ absorption in the models with [Fe/H] = $-2$ and $-2.5$ leads to a negative $\Delta_{\rm NLTE}$. At the same time, the core strengthening
for Ca~I 6439\,\AA\ determines the negative sign of $\Delta_{\rm NLTE}$ even at solar metallicity, and the effect is enhanced as the wings weaken in the models with [Fe/H] = $-1$ and $-2$. At lower metallicity all the Ca~I lines, except the 4226\,\AA\ resonance line, are formed in the
layers where overionization dominates, and $\Delta_{\rm NLTE}$ is everywhere positive.

\section{INTERPOLATION OF THE NON-LTE CORRECTIONS}

Our non-LTE calculations were made in the following ranges of stellar parameters: 4000~K $\le \Teff \le$ 5000~K with a 250~K step, 0.5 $\le$ log~g $\le$ 2.5 with a 0.5 step, and $0 \le$ [Fe/H] $\le -4$ with a variable step, 0.5~dex at [Fe/H] $\le -2$ and 1~dex at a higher metallicity. Everywhere, $\xi_t$ = 2\,\kms. The tables include the non-LTE corrections for 28/42/54/262 lines of Ca~I/Ti~I/Ti~II/Fe~I and the corresponding equivalent widths calculated in the LTE approximation. The data are accessible at the site {\tt http://spectrum.inasan.ru/nLTE/}, and the
non-LTE correction for the spectral line being studied can be obtained online by interpolation for given atmospheric parameters.

The Web interface (Fig.\,\ref{Fig:online}) is a form that consists of a dropdown list of elements (Ca~I, Ti~I, Ti~II, and Fe~I) and four text fields designed to enter a spectral line wavelength (in \,\AA), an effective temperature $\Teff$, a surface gravity log~g, and an elemental abundance
[X/H] by the user. The ranges of admissible values are specified above each field. Since the data are presented for a specific list of lines, their correct identification between the list and the wavelength specified by the user is important. For this purpose, a prompt is organized in the entry field. When the first digits of a wavelength are entered, a list of wavelengths starting
with these digits is displayed, and it is possible to choose the needed value. When the values outside the grid used are entered, an error message is generated.
In the case of a successful entry, the parameters are transferred to the code on the server. The code is written in PDL (Perl data language) and implements a trilinear interpolation of the non-LTE corrections and equivalent widths. The result is displayed without reloading the entire page.

For most lines in the table there are grid points $\Teff$/log~g/[X/H] with missing data of corrections. This means that the line is very weak, with $EW \le$ 3~m\AA\ at given parameters. If the data are absent at least at one of the grid points, then no interpolation is made,
and the 'I cannot interpolate' is generated.

\bigskip
{\it Methodical recommendations}
\begin{enumerate}
\item For the models with [Fe/H] $\le -1$ the non-LTE calculations for Ca~I and Ti~I-Ti~II were performed with an abundance increased by 0.4~dex compared to the iron abundance, i.e., with [X/Fe] = 0.4. Therefore, when interpolating the corrections, we should specify ([X/H] - a) but not the stellar metallicity. Here, a = 0.4 for the Ca~I, Ti~I, and Ti~II lines in MP stars and a = 0.0 in all the remaining cases.
\item If the abundance derived after adding the non-LTE correction differs by more than 0.15~dex from [X/H] specified during the interpolation, then the interpolation procedure should be repeated with the new value of [X/H].
\item The interpolation code outputs not only $\Delta_{\rm NLTE}$, but also $EW$(LTE) corresponding to the specified Teff/log~g/[X/H]. We recommend to check $EW$ by
comparing it with the theoretical value from the user's calculations or the observed one.
\end{enumerate}

{\it Acknowledgements.}
This work was supported in part by the Basic Research Program P-7 of the Presidium of the
Russian Academy of Sciences. L. Mashonkina and T. Sitnova are grateful to the International Space Science Institute (ISSI) in Bern (Switzerland) for supporting and financing the
'First Stars in Dwarf Galaxies' and 'Formation and Evolution of the Galactic Halo' International Teams.

\end{document}